\documentstyle[pra,aps,graphicx]{revtex}

\newcommand{\OO}{\mbox{\boldmath $\Omega$}}

\newcommand{\ka}{\mbox{\boldmath $\kappa$}}
\newcommand{\n}{\mbox{\boldmath $n$}}

\newcommand{\vv}{\mbox{\boldmath $v$}}
\newcommand{\rr}{\mbox{\boldmath $r$}}
\newcommand{\q}{\mbox{\boldmath $q$}}
\newcommand{\PP}{\mbox{\boldmath $P$}}
\newcommand{\p}{\mbox{\boldmath $p$}}
\newcommand{\kk}{\mbox{\boldmath $k$}}
\newcommand{\ro}{\mbox{\boldmath $\rho$}}
\begin{document}
\draft
\title
{Contradictions of quantum scattering theory}
\author
{Vladimir K. Ignatovich }
\address{Frank Laboratory of Neutron Physics of Joint Institute for Nuclear
Research, \\141980, Dubna Moscow region, Russia}

\date{\today}
\maketitle

\begin{abstract}
The standard scattering theory (SST) in non relativistic quantum
mechanics is analyzed. Contradictions of SST are revealed.
Justification of SST in textbooks with the help of fundamental
scattering theory (FST) is shown to be unconvincing. A theory of
wave packets scattering on a fixed center is presented, and its
similarity to plane wave scattering is demonstrated. The neutron
scattering in a monatomic gas is investigated, and ambiguity of
the cross section is found. The ways to resolve contradictions and
ambiguity is suggested, and experiments to explore properties of
wave packets are discussed.
\end{abstract}

\section{Introduction}

At present we have three scattering theories:
\begin{enumerate}
\item
Theory of spherical harmonics (SHT) used for elastic scattering
(see, for example~\cite{land});
\item
More general theory of Van Hove correlation functions, which we
call standard scattering theory (SST) (see, for
example~\cite{ml,lov});
\item
The fundamental scattering theory (FST) (see, for
example~\cite{gbw,tay}), which is used mainly for justification of
SST.
\end{enumerate}
The first one is inconsistent, the second one is not a theory but
only a set of rules for calculation of cross sections, and the
third one contains an error, so it does not justify SST. The error
is hidden in definition of scattering probability. It relates
initial wave packet of the incident neutron to outgoing plane wave
of the scattered neutron. Such a definition completely violates
unitarity. The unitarity is conserved when both initial and final
particles are plane waves or wave packet, but in both cases FST
gives not cross sections but only probabilities.

We applied the fundamental theory of plane waves scattering
(FPWST) (without generally used artificial finite volume $L^3$) to
neutron scattering in monatomic gas, and found that the result is
ambiguous, which is very easily checked.

The probability amplitude $dF(\kk_i,\p_i\to\kk_f,\p_f)$ of the
neutron scattering on an atom is (see section 4)
\begin{equation}\label{intr}
dF(\kk_i,\p_i\to\kk_f,\p_f)=
d^3k_fd^3p_f\frac{ib}{\pi}\delta(\kk_i+\p_i-\kk_f-\p_f)\delta(k_f^2+\mu
p_f^2-k_i^2-\mu p_i^2),
\end{equation}
where $\kk_{i,f}$ are neutron's and $\p_{i,f}$ are atom's initial
and final momenta, $\mu=m/M$ is ratio of the neutron's $m$ to the
atomic $M$ masses. The amplitude (\ref{intr}) is the product of
scattering amplitude $b$ and two $\delta$-functions related to the
momentum and energy conservation laws. It is simplified after
integration over $d^3p_f$, which eliminates the first
$\delta$-function, and is reduced to
\begin{equation}\label{intr1}
dF(\kk_i\to\kk_f,\p_i)=
d^3k_f\frac{ib}{\pi}\delta(k_f^2+\mu(\kk_i+\p_i-\kk_f)^2-k_i^2-\mu
p_i^2)\equiv d^3k_f\frac{ib}{2\pi}\delta(E_R-\omega+\mu\p_i\ka),
\end{equation}
where $\ka=\kk_i-\kk_f$, $\omega=k_i^2/2-k_f^2/2$, and
$E_R=\mu\kappa^2/2$.

Expressions (\ref{intr}) and (\ref{intr1}) are very natural,
however the two ways of probability calcu\-la\-ti\-on: one
directly in laboratory reference frame (LF), and another one via
center of mass (CM) reference frame give two different results.

\subsection{Short explanation}

\paragraph {In LF} we directly integrate over $dk_f$
and get
$dF(\kk_i,\Omega_f,\p_i)=F'(\kk_i,\Omega_f,\p_i)d\Omega_f$, where
$\Omega_f$ is neutron scattering angle, and
$F'(\kk_i,\Omega_f,\p_i)=dF(\kk_i,\Omega_f,\p_i)/d\Omega_f$. With
$F'(\kk_i,\Omega_f,\p_i)$ the probability
$dw_{l}(\kk_i,\Omega_f,\p_i)$ of scattering in LF is
$dw_{l}(\kk_i,\Omega_f,\p_i)=|F'(\kk_i,\Omega_f,\p_i)|^2d\Omega_f$.

\paragraph{In the standard approach} (look, for instance~\cite{lov})
we transform (\ref{intr1}) to CM, i.e. change variable:
$\kk_f=\kk_c+\mu(\kk_i+\p_i)/(1+\mu)$, integrate over $dk_c$, find
$dF(\kk_i,\Omega_c,\p_i)=F'_c(\kk_i,\Omega_c,\p_i)d\Omega_c$ of
neutron scattering into the CM scattering angle $\Omega_c$, and
obtain pro\-ba\-bi\-li\-ty
$dw_c(\kk_i,\Omega_c,\p_i)=|F'_c(\kk_i,\Omega_c,\p_i)|^2d\Omega_c$.
Transformation of this probability back to LF gives $d\tilde
w_l(\kk_i,\Omega_f,\p_i)$.

\paragraph{The ambiguity is:}$d\tilde
w_l(\kk_i,\Omega_f,\p_i)\ne dw_l(\kk_i,\Omega_f,\p_i)$.

\subsection{Details of calculations} Now we present these
calculations in details, but first we want to stress, that the
direct calculation, which is the main new point, does not include
change of reference frames, while all the calculations with change
of frames lead to the standard result. So it is not profitable to
seek for errors in change of reference frames.

\subsubsection{Direct calculation in LF}

Since $d^3k_f=k^2_fdk_fd\Omega_f$, we directly integrate
$\delta$-function in (\ref{intr1}) over $dk_f$ and get
\begin{equation}\label{intr2}
d^3k_f\delta(k_f^2+\mu(\kk_i+\p_i-\kk_f)^2-k_i^2-\mu p_i^2)=
d\Omega_f\frac{k_f^2}{2|(1+\mu)k_f-\mu(\n\PP)|},
\end{equation}
where $\PP=\kk_i+\p_i$ is the total momentum of the neutron and
atom,
\begin{equation}
k_f=\frac{\mu\PP\n\pm\sqrt{\mu^2(\PP\n)^2-\mu^2P^2+(\kk_i-\mu\p_i)^2}}
{1+\mu}>0, \label{intr3}
\end{equation}
and $\n$ is a unit vector in the scattering direction of $\OO_f$.
Substitution of (\ref{intr2}) into (\ref{intr1}) gives
\begin{equation}\label{intr4}
dF(\kk_i,\Omega_f,\p_i)=d\Omega_f\frac{ib}{2\pi}\frac{k_f^2}
{\sqrt{\mu^2(\PP\n)^2-\mu^2P^2+(\kk_i-\mu\p_i)^2}},
\end{equation}
and the probability of scattering into $d\Omega_f$ is
\begin{equation}\label{intr5}
dw(\kk_i,\Omega_f,\p_i)=\left|\frac{dF(\kk_i,\p_i,\Omega_f)}{d\Omega_f}\right|^2d\Omega_f=
\left|\frac{ib}{2\pi}\right|^2\frac{k_f^4d\Omega_f}
{|\sqrt{\mu^2(\PP\n)^2-\mu^2P^2+(\kk_i-\mu\p_i)^2}|^2}.
\end{equation}
After backward transformation (\ref{intr2}) one restores
$\delta$-function of energy conservation and obtains
$$dw(\kk_i\to\kk_f,\p_i)=$$
\begin{equation}\label{intr6}
\left|\frac{ib}{2\pi}\right|^2\frac{2k_f^2d^3k_f\delta(k_f^2+\mu
p_f^2-k_i^2-\mu p_i^2)}
{\sqrt{\mu^2(\PP\n)^2-\mu^2P^2+(\kk_i-\mu\p_i)^2}}= d^3k_fk_f^3
\left|\frac{b}{2\pi}\right|^2 \frac{\delta(E_R-\omega+\mu\ka\p_i)}
{|s-\mu\omega-\mu\kk_i\p_i|},
\end{equation}
where $s=k_i^2/2+k_f^2/2$, $\n k_f$ was replaced with $\kk_f$, and
we used relation
\begin{equation}\label{intr1a}
2\delta(k_f^2+\mu(\kk_i+\p_i-\kk_f)^2-k_i^2-\mu p_i^2)=
\delta(E_R-\omega+\mu\p_i\ka).
\end{equation}
We see, that the scattering probability (it is obtained directly
in the LF, without change to other reference frames) depends on 3
variables $\kappa$, $\omega$ and $s$ contrary to the wide spread
belief that it depends only on two variables $\kappa$ and
$\omega$.

\subsubsection{Calculation via CM reference frame}

The argument of the $\delta$-function can be represented as
\begin{equation}
k_f^2+\mu(\kk_i+\p_i-\kk_f)^2-k_i^2-\mu p_i^2=
(1+\mu)\left(\kk_f-\frac{\mu}{1+\mu}\PP\right)^2-
\frac{(\kk_i-\mu\p_i)^2}{1+\mu}, \label{intr7}
\end{equation}
where $\PP=\kk_i+\p_i$ is the total momentum of the CM, and
$\kk_i-\mu\p_i$, later denoted by $\q$, is the relative speed of
the neutron and atom. After change of variables
\begin{equation}
\kk_{c}=\kk_f-\mu\PP/(1+\mu), \label{intr8}
\end{equation}
we obtain
\begin{equation}
\delta(k_f^2+\mu(\kk_i+\p_i-\kk_f)^2-k_i^2-\mu p_i^2)d^3k_f=
\delta\left((1+\mu)k_c^2-q^2/(1+\mu)\right)d^3k_c, \label{intr7a}
\end{equation}
Integration of the $\delta$-function over $dk_{c}$ gives
\begin{equation}
\delta(k_f^2+\mu(\kk_i+\p_i-\kk_f)^2-k_i^2-\mu
p_i^2)d^3k_f=\delta\left[(1+\mu)k_{c}^2-
\frac{\q^2}{1+\mu}\right]d^3k_c= \frac{qd\Omega_c}{2(1+\mu)^2},
\label{intr7b}
\end{equation}
and (\ref{intr1}) is reduced to
\begin{equation}\label{intr9}
dF_c(\kk_i,\Omega_c,\p_i)=\frac{ibq}{2\pi(1+\mu)^2} d\Omega_{c}.
\end{equation}
The scattering probability from an atom with momentum $\p_i$ is
\begin{equation}\label{intr10}
dw_c(\kk_i,\Omega_c,\p_i)=\left|\frac{ibq}{2\pi(1+\mu)^2}\right|^2
d\Omega_{c}.
\end{equation}
The backward transformation (\ref{intr7b}) restores
$\delta$-function
\begin{equation}\label{intr11a}
d\tilde
w_l(\kk_i\to\kk_f,\p_i)=\left|\frac{ib}{2\pi(1+\mu)}\right|^22d^3k_f|\kk_i-\mu\p_i|
\delta(k_f^2+\mu(\kk_i+\p_i-\kk_f)^2-k_i^2-\mu p_i^2),
\end{equation}
and with account of (\ref{intr1a}) we get
\begin{equation}\label{intr11m}
d\tilde
w_l(\kk_i\to\kk_f,\p_i)=\left|\frac{ib}{2\pi(1+\mu)}\right|^2d^3k_f
|\kk_i-\mu\p_i|\delta(E_R-\omega+\mu\p_i\ka).
\end{equation}
We want to point out, that the last expression after omission of
the factor $|\kk_i-\mu\p_i|^2/(2\pi)^2$, leads to the standard
cross section
\begin{equation}\label{in11m}
d\sigma_l(\kk_i\to\kk_f,\p_i)=\frac{|b|^2}{(1+\mu)^2}\frac{d^3k_f}{
|\kk_i-\mu\p_i|}\delta(E_R-\omega+\mu\p_i\ka),
\end{equation}
which proves that all changes of reference frames were performed
properly.

After integration of (\ref{intr11a}) over $dk_f$ we obtain
\begin{equation}\label{intr12}
d\tilde
w_l(\kk_i,\Omega_f,\p_i)=\left|\frac{ib}{2\pi(1+\mu)}\right|^2
\frac{k_f^2|\kk_i-\mu\p_i|d\Omega_f}{\sqrt{\mu^2(\PP\n)^2-\mu^2P^2+(\kk_i-\mu\p_i)^2}}.
\end{equation}
Both (\ref{intr11m}), (\ref{intr6}) and (\ref{intr12}),
(\ref{intr5}) are different.

\subsection{The content of the paper}

The FPWST, if we accept it notwithstanding of the ambiguity (it
will be later eliminated), gives probabilities and not cross
sections. To see how to match them to experiment, we in the next
section discuss what do experimentalists really measure, and how
to get a cross section from probability. This discussion
inevitably leads to a wave packet description of free neutron.

In the third section we show what are the drawbacks of all the
three theories pointed above, in fourth section we derive
scattering amplitude (\ref{intr}) and demonstrate how to get the
standard cross section for neutron scattering in monatomic gas.

In 5-th section some wave packets and their properties are
discussed. In particular, we consider scattering of a wave packet
from a fixed center and show that in linear theory probability of
scattering does not depend on impact parameter, so to get a cross
section from probability we need a nonlinear interaction.

In 6-th section some ways to resolve contradictions and
ambiguities are proposed, in section 7 some experiments to
investigate wave packet properties are discussed, and in
conclusion the results of the paper are summarized.

\section{What is the scattering cross section}

Almost all experiments (exceptions are reflectometry and
diffractometry) are interpreted in terms of scattering cross
sections. Here we analyze what is really measured, how cross
section is extracted and how it is theoretically defined. This
analysis leads to conclusion that to get a cross section from
probability we have to introduce a parameter $A$ with dimension of
area, characterizing the size of the neutron wave function.

\subsection{Definition of the scattering cross section in an
experiment}

\begin{figure}[t]
{\par\centering\resizebox*{8cm}{!}{\includegraphics{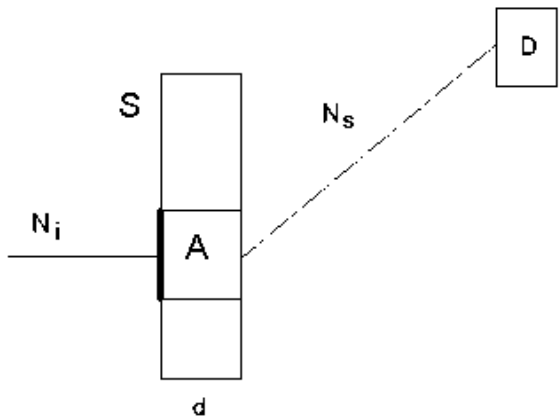}}\par}
\caption{Definition of the cross section for scattering on a
single atom} \label{f1}
\end{figure}

For definition of the scattering cross section we can look at an
experiment schematically shown in fig. \ref{f1}. If the detector
registers $N_s$ scattered neutrons per unit time, then the total
probability $W(\kk_i\to\kk_f)$ for a single neutron to be
scattered in the sample into a given direction is
\begin{equation}\label{prob}
W(\kk_i\to\kk_f)=\frac{N_s}{N_i}=\frac{N_s}{JS},
\end{equation}
where $J$ is the neutron flux density, $S$ is the area of the sample immersed into the neutron flux, and
$N_i=JS$ is the total number of neutrons incident on the sample per unit time.

Experimentalists divide this value by dimensional parameter
$N_0d$, where $N_0$ is atomic number density in the sample, and
$d$ is the sample width. As a result one obtains the cross
section\footnote{We assume the sample to be thin for not to take
into account self-shielding.}
\begin{equation}\label{ratio1}
\sigma(\kk_i\to\kk_f)=\frac{N_s}{N_iN_0d}=\frac{N_s}{JSN_0d}=\frac{N_s}{JV},
\end{equation}
where $V=Sd$ is the volume of the sample illuminated by the
incident neutron flux. The expression (\ref{ratio1}) is commonly
accepted, but gives no insight into interaction of a single
neutron with a single atom.

It looks more reasonable from total probability $W(\kk_i\to\kk_f)$
(\ref{prob}) of scattering of a single neutron in the whole sample
to find the scattering probability $w_1$ of a single neutron on a
single atom:
\begin{equation}\label{prob1}
w_1(\kk_i\to\kk_f)=\frac{W(\kk_i\to\kk_f)}{N_a},
\end{equation}
where $N_a$ is the number of atoms met by a single neutron on its
way during the flight through the sample. To find the number of
atoms on the neutron's way we have to introduce a front area $A$
of the incident particle wave function, and suppose that
scattering takes place only, if the scattering center crosses this
area. So, let the neutron wave function to have area $A$, then
$N_a=N_0Ad$. From (\ref{prob1}) we immediately find the scattering
cross section of a single neutron per single atom:
\begin{equation}\label{prob2}
\sigma(\kk_i\to\kk_f)=Aw_1(\kk_i\to\kk_f)=\frac{W(\kk_i\to\kk_f)}{N_0d}=
\frac{N_s}{N_iN_0d}=\frac{N_s}{JSN_0d}=\frac{N_s}{JN_0V},
\end{equation}
which coincides with (\ref{ratio1}). The left hand side is the
cross section we must calculate, the right hand side is the
experimentally defined cross section. To compare theory with
experiment we must be able to calculate $w_1(\kk_i\to\kk_f)$ and
$A$.

\subsection{Phenomenological definition of the scattering cross section}

According to all the textbooks the scattering cross section is
defined as a ratio of the count rate $N_s$ of scattered particles
to the flux density, $J$, of the incident particles:
\begin{equation}\label{ratio}
\sigma=\frac{N_s}{J}.
\end{equation}
If, for instance, we have a small sample, the above ratio gives
the cross section of the whole sample, and if we divide this ratio
by the total number of atoms in the sample, $N_a=N_0V$,
illuminated by the incident flux, we obtain the result
(\ref{ratio1}), which defines the cross section per one atom. We
call it a phenomenological definition, because it says nothing
about interaction of a single neutron with a single atom.

\subsection{Theoretical definition of the scattering cross section}

Theoretically, if you want to find a number of scattered particles
for a single target atom and a given incident flux $J$, you must
first find, how a single particle is scattered, and how this
scattering depends on impact parameter. After that we should
integrate over all particles in the incident flux, and average
over all possible positions of the scatterer. This procedure gives
the number $N_s$ of scattered particles for the given $J$. The
ratio $N_s/J$ is an average cross section $Aw_1$, and it can be
compared with the phenomenological one.

The parameter $A$ includes also dimension of a single nucleus. We
can even suppose that it is related only to nucleus, and neutrons
are point particles, propagating like rays of a wave. In that case
$A$ is equal to the size $\sigma_N$ of the nucleus, and the total
cross section can never be larger than $\sigma_N$, because the
total probability $w_1$ can never be larger than unity. This
contradicts to the well known facts, that some capture cross
sections can be many orders of magnitude larger than $\sigma_N$.
To avoid this contradiction, we have to assume that $A$ is related
to neutron and is considerably larger than $\sigma_N$, so we can
neglect contribution of $\sigma_N$ into $A$.

\subsection{On wave packets}

Introduction of the finite front area means that the particle wave
function is not a plane wave, but a wave packet. This wave packet
cannot be spreading, because, if it were, the transmission of a
sample would decrease, when the sample is shifted from source to
detector, and no one, so far, had ever observed such a phenomenon.

One of the possible candidates for the nonspreading wave packet is
the singular de Broglie wave packet (dBWP)~\cite{bro,uts,uts1}
\begin{equation}
\psi_{dB}(\rr,\vv,t,s)=
\sqrt{\frac{s}{2\pi}}\exp(i\kk\rr-i\omega(k,s)
t)\frac{\exp(-s|\rr-\vv t|)} {|\rr-\vv t|}, \label{d1}
\end{equation}
where $\omega(k,s)=[k^2-s^2]/2$, $s$ determines the packet width,
and $\vv$ is the wave packet velocity, which in our units
$m=\hbar=1$ coincides with wave vector $\kk$. The front area of
(\ref{d1}) can be estimated as $A_{dB}=\pi/s^2$. Some reasonings
related to solution of ultra cold neutrons anomaly~\cite{uts},
leads to
\begin{equation}\label{area1}
s\approx3\cdot10^{-5}k,
\end{equation}
which means that for thermal energies $A_{dB}\gg\sigma_N$. Because
of that the short range interaction becomes a long range one,
which is manifested in such effects as total reflection and
diffraction in crystals. For description of these processes one
does not use cross sections, so the parameter $A$ can be even
omitted, and neutrons can be represented by plane waves.

\section{Three scattering theories}

At present we have three theories and every one has its own
drawbacks.
\begin{description}
\item
[SHT.] It is used only for elastic scattering and two particles
scattering in CM reference frame. It is logically inconsistent and
contradicts to canonical quantum mechanics (CQM), because it does
not describe free particles after scattering. The spherical waves
are not solutions of the free Schr\"odinger equation. To make SHT
self consistent one have to find asymptotical form of spherical
waves, which is a superposition of free states or free plane
waves. However after this step no cross section appears in theory,
but only dimensionless probabilities.
\item
[SST.] This theory, which uses Van Hove correlation functions, is
not a theory, but only a set of rules how to calculate a cross
section. This ``theory'' uses free states after scattering,
however it never calculates scattering amplitudes, operates with
proba\-bi\-li\-ty per unit time, which does not depend on time,
introduces superfluous space cell of arbitrary large size $L$, a
flux density for a single particle and has other drawbacks
discussed below.
\item
[FST.] Its main goal is to justify SST. It uses asymptotical wave
function after scattering, which is a superposition of free
states. More over it calculates scattering probabilities. However
these probabilities violate unitarity, and transition from
probabilities to cross sections contains some connivance, which is
not valid.
\end{description}
Now we shall consider them in details

\subsection{Theory of spherical harmonics (SHT)}

In this theory the wave function of a scattered neutron is
represented (see, for example,~\cite{land}) as
\begin{equation}\label{fq}
\Psi=\exp(i\kk\rr)-\frac{f(\vartheta)}{r}\exp(ikr),
\end{equation}
where plane $\exp(i\kk\rr)$ and spherical $\exp(ikr)/r$ waves describe
incident and scattered particles, respectively. The
scattering amplitude is
\begin{equation}\label{fq1}
f(\vartheta)=\frac1{2ik}\sum\limits_{l=0}^\infty(2l+1)(S_l-1)P_l(\cos\vartheta)
\end{equation}
where $S_l$ are scattering matrix elements of spherical harmonics
and $P_l(\cos\vartheta)$ are Legeandre polynomials.

The simplest process is elastic s-wave scattering from a fixed
center with wave function,
\begin{equation}\label{fq2}
\Psi=\exp(i\kk\rr)-\frac{b}{r}\exp(ikr),
\end{equation}
in which $f(\vartheta)=b=$const.

The approach with spherical waves is borrowed from classical
theories of sound and classical light scattering, but it is not a
quantum scattering theory, because the spherical wave does not
describe a free particle. It satisfies the equation
\begin{equation}
[\Delta+k^2]\frac{\exp(ikr)}{r}=-4\pi\delta(\rr),
\label{q2a}
\end{equation}
with the right hand side containing the Dirac $\delta$-function.
It is not identical zero in the whole space. Therefore (\ref{q2a})
is not a free equation. Usually one says that we need wave
function outside the point $r=0$, and there the spherical wave
satisfies the free Schr\"odinger equation. However even the wave
function of the bound state does satisfy free Schr\"odinger
equation outside the potential. Nevertheless we do not consider
the tails of the bound wave functions as free particles. Of
course, in the case of bound states the wave function tails decay
exponentially at infinity, and this fact justifies why these tails
are not considered as free particles. We shall show below, that
the asymptotic $1/r$ of spherical waves also contains
exponentially decaying waves. Therefore to get real asymptotic of
spherical waves at infinity we need to exclude exponentially
decaying part from it too (compare (\ref{2a}) and (\ref{2ad})
below).

To make SHT really quantum theory we have to find asymptotical
form of spherical waves, when particles are sufficiently separated
after scattering. This asymptotical wave function should be a
superposition
\begin{equation}\label{fqq2}
\Psi\Rightarrow\exp(i\kk\rr)-\int d\Omega
F'(k,\Omega)\exp(i\kk_\Omega\rr),
\end{equation}
where $\kk_\Omega$ is wave vector of the length
$|\kk_\Omega|=|\kk|$ pointing into the direction $\OO$ in the
element $d\Omega$ of the solid angle $\Omega$. Then the
probability of scattering into $d\Omega$ is
\begin{equation}\label{fqq3}
dw(k,\Omega)=|F'(k,\Omega)|^2d\Omega.
\end{equation}

The asymptotic wave function (\ref{fqq2}) can be found in two
ways: stationary and non\-stati\-o\-na\-ry ones.

\subsubsection{The stationary way}

The spherical function has 2-dimensional Fourier expansion
\begin{equation}
\frac{\exp(ikr)}{r}=\frac{i}{2\pi}\int\exp(i\p_\|\rr_\|+ip_z|z|)
\frac{d^2p_\|}{p_z},
\label{2a}
\end{equation}
where we fix the direction from the scatterer to the observation
point as $z$-axis, and integrate over all components $\p_\|$
parallel to $x,y$ plane with $z$-component of the momentum being
equal to $p_z=\sqrt{k^2-p_\|^2}$.

The range of integration over $\p_\|$ (\ref{2a}) is infinite, and,
in particular, it includes those $\p_\|$, for which $p_\|^2>k^2$.
At these $\p_\|$ the component $p_z$ is imaginary, and
$\exp(ip_z|z|)$ is an exponentially decaying function. If the
distance to the observation point is large enough (later we
discuss what does it mean ``enough''), we can neglect
exponentially decaying terms, and restrict integration to
$p_\|^2\le k^2$:
\begin{equation}
\frac{\exp(ikr)}{r}\approx\frac{i}{2\pi}\int\limits_{p_\|^2<k^2}
\exp(i\p_\|\rr+ip_z|z|)
\frac{d^2p_\|}{p_z}.
\label{2ad}
\end{equation}
In this integral we can substitute
\begin{equation}
\frac{d^2p_\|}{p_z}=d^3p\delta(p^2/2-k^2/2)\Theta(p_zz>0),
\label{sph8}
\end{equation}
where $p^2=p_\|^2+p_z^2$, $p_z$ is a variable, and we introduced
the step function $\Theta(x)$, which is unity or zero, when inequality
in its argument is satisfied or not, respectively. Substitution
of (\ref{sph8}) into (\ref{2ad}) gives
\begin{equation}
\frac{\exp(ikr)}{r}\approx
\frac{i}{2\pi}\int\limits_{}^{}\exp(i\p\rr)\Theta(p_zz>0)
\,d^3p\delta(p^2/2-k^2/2)=
\frac{ik}{2\pi}\int\limits_{4\pi}^{}\exp(i\kk_\Omega\rr)d\Omega,
\label{sph5}
\end{equation}

The terms which we neglected are of the order $1/r$ because
at the observation point $z$ ($\rr_\|=0$) we have
\begin{equation}
\frac{1}{2\pi}\left|\int\limits_{p_\|^2>k^2}
\exp\left(-\sqrt{p_\|^2-k^2}z\right)\frac{d^2p_\|}{\sqrt{p_\|^2-k^2}}\right|=
\frac{1}{r},
\label{2ae}
\end{equation}
where we replaced $z$ by the distance $r$ between scatterer and observation point.

Thus the asymptotical form of the wave function (\ref{fq2}) after scattering is just (\ref{fqq2})
with scattering probability amplitude
\begin{equation}
F'(k,\Omega)=\frac{ibk}{2\pi}=i\frac{b}{\lambda}, \label{sph7}
\end{equation}
and scattering probability
\begin{equation}
dw(k,\Omega)=|F'(k,\Omega)|^2d\Omega=\left|\frac{b}{\lambda}\right|^2d\Omega,\quad
w(k)=\int\limits_{4\pi}^{}dw(k,\Omega)=4\pi
\left|\frac{b}{\lambda}\right|^2, \label{sph7a}
\end{equation}
where $\lambda=2\pi/k$ is the neutron wave length. We see that
(\ref{fq2}) is reduced to (\ref{fqq2}), when we neglect the terms
of the order $b/r$. Since the decision to neglect or not to
neglect this term is at will of the physicist, then the distance
$r$ from the center is not an asymptotical one, being even of
light years size, if he uses the spherical wave. On the other side
the distances of the order 1 \AA\ are asymptotical ones, if $b/r$
is neglected.

\subsubsection{The nonstationary derivation of asymptotical wave function
at large times $t\to\infty$}

To find nonstationary asymptotic of the wave function (\ref{fq2})
we include in it the time dependent factor $\exp(-i\omega_kt)$,
where $\omega_k=k^2/2$, and use 3-dimensional Fourier expansion
for the spherical wave
\begin{equation}
\delta\psi(r,t)=\frac{b}{r}\exp(ikr-i\omega_kt)=\frac{b}{(2\pi)^2}\int\limits_{}^{}
\frac{d^3p}{\omega_p-\omega_k-i\epsilon}\exp(i\p\rr-
i\omega_kt),\qquad \quad\omega_p=p^2/2. \label{a}
\end{equation}

One adds and subtracts $i\omega_pt$ in the exponent, and
represents the field (\ref{a}) as a superposition of plane waves
\begin{equation}
\delta\psi=\int\limits_{}^{}\widetilde{F}'(\p,t)\exp(i\p\rr-i\omega_pt)d^3p,
\label{mt3}
\end{equation}
with amplitudes
\begin{equation}
\widetilde{F}'(\p,t)=\frac{b}{(2\pi)^2}
\frac{\exp(i[\omega_p-\omega_k]t)}{\omega_p-\omega_k-i\epsilon},
\label{ps8}
\end{equation}
which depend on time $t$.

There is a relation~\cite{gbw}
\begin{equation}
\frac{\exp(i[\omega_p-\omega_k]t)}{\omega_p-\omega_k-i\epsilon}
=i\int\limits_{-\infty}^{t}\exp(i[\omega_p-\omega_k]t')dt',
\label{mt1}
\end{equation}
which in the limit $t\to\infty$
gives the law of energy conservation:
\begin{equation}
i\lim\limits_{t\to\infty}\int\limits_{-\infty}^{t}\exp(i[\omega_p-\omega_k]t')
dt'=2\pi i\delta(\omega_p-\omega_k)=4\pi i\delta(p^2-k^2).
\label{mt2}
\end{equation}
In this limit (\ref{mt3}) is
\begin{equation}
\delta\psi=
\int\limits_{}^{}\frac{ib}{\pi}\exp(i\p\rr-i\omega_pt)
d^3p\delta(p^2-k^2)=
\frac{ibk}{2\pi}
\int\limits_{4\pi}^{}d\Omega\exp(i\kk_\Omega\rr-i\omega_kt),
\label{mt4}
\end{equation}
and we get dimensionless scattering probability amplitude
(\ref{sph7}) and the total scattering probability
$w(k)=4\pi|b/\lambda|^2$, which coincide with (\ref{sph7a}).

The main result of all these considerations is: the correct
asymptotical wave function gives not a cross section, but only
dimensionless probability.

\subsubsection{Phenomenological definition of cross section}

With asymptotical wave function (\ref{mt4}) it is possible to
define phenomenological semi integral cross section $\sigma(\n)$
as a ratio of normal flux $J(\n)$ through an infinite plane
$S(\n)$ with normal $\n$ to the incident flux density $k$
\begin{equation}
\sigma(\n)=\frac1{2ik}\int
d^2r_\parallel\left[\delta\psi^*(\rr)\overrightarrow{\frac{d}{dz}}\delta\psi(\rr)
-\delta\psi^*(\rr)\overleftarrow{\frac{d}{dz}}\delta\psi(\rr)\right].
\label{mmt4}
\end{equation}
The wave function (\ref{mt4}) in general case, when scattering is
nonisotropic ($b=b(\kk_\Omega)$), can be represented as
\begin{equation}
\delta\psi(\rr)= \frac{ik}{2\pi} \int\limits_{4\pi}^{}d\Omega
b(\kk_\Omega) \exp(i\kk_\Omega\rr-i\omega_kt)=\frac{i}{2\pi}
\int\limits_{p_\parallel<k}^{}\frac{d^2p_\parallel}{p_\perp} b(\p)
\exp(i\p\rr-i\omega_kt), \label{mmt3}
\end{equation}
where $p_\perp=\sqrt{k^2-p^2_\parallel}$. Substitution of this
function into (\ref{mmt4}) gives
\begin{equation}
\sigma(\n)=\int\limits_{p_\parallel<k}^{}\frac{d^2p_\parallel}{kp_\perp}
|b(\p)|^2=\int\limits_{2\pi(\n)} |b(\kk_\Omega)|^2d\Omega,
\label{mmt2}
\end{equation}
where $2\pi(\n)$ denotes solid angle $2\pi$ around normal vector
$\n$. We can define differential cross section as
\begin{equation}
\frac{d\sigma}{d\Omega}=\frac{d\sigma(\n)}{d\n}=|b(\kk_\Omega)|^2.
\label{mmt1}
\end{equation}
This definition is not unique, because some arbitrary function
$F(\Omega)$ can be added to it, for which
$$\int\limits_{2\pi}F(\Omega)d\Omega=0,$$
and we do not want to accept it, because it is a phenomenological
one.

\subsection{The Standard Scattering Theory (SST)}

SST (see, for example~\cite{ml,lov}) is more general than SHT,
because it is applied to both elastic and inelastic processes. It
lists the rules, to be justified by FST, one must use to derive
scattering cross section. These rules are:
\begin{enumerate}
\item
Define the cross section as a ratio
\begin{equation}\label{nqb4}
d\sigma=\frac1{J_i}dW_F,
\end{equation}
where $J_i$ is the flux density of incident neutrons, and $dW_F$
is probability of scattering per unit time:
\item
Define the probability of scattering per unit time according to the Fermi Golden Rule
\begin{equation}\label{wpg1a}
dW_F(\kk_i,\lambda_i\to\kk_f,\lambda_f,t)=\frac{2\pi}{\hbar}\left|\langle\lambda_f,\kk_f|V|\lambda_i,
\kk_i\rangle\right|^2\rho(E_{fk}),
\end{equation}
where $\langle\lambda_f,\kk_f|V|\lambda_i,
\kk_i\rangle$ is a matrix element of the neutron-scatterer interaction
potential $V$ between initial $|\kk_i>$, $|\lambda_i>$, and final $|\kk_f>$,
$|\lambda_f>$,
neutron and scatterer states, respectively, and $\rho(E_{fk})$ is the density of the
neutron final states:
\begin{equation}
\rho(E_{fk})=\delta(E_{ik}+E_{i\lambda}-E_{fk}+E_{f\lambda})\frac{L^3d^3k_f}{(2\pi)^3}.
\label{wpg1b}
\end{equation}
The delta-function factor corresponds to the energy conservation
law. It contains initial and final neutron $E_{i,fk}$ and
scatterer $E_{i,f\lambda}$ energies. The last factor is the phase
space density of the neutron final states, which includes momentum
element $d^3k_f$ and some volume $L^3$ with an arbitrary large
size $L$.
\item
Define neutron states before and after scattering as (they are
really free states)
\begin{equation}\label{wp1b}
|\kk_i>=L^{-3/2}\exp(i\kk_{i,f}\rr),
\end{equation}
and with them the flux density of the incident neutron
\begin{equation}\label{wp2b}
J_i=\hbar \frac{k_i}{L^3}.
\end{equation}
\item
The ratio (\ref{nqb4}) gives the cross section
\begin{equation}\label{g1a}
d\sigma(\kk_i,\lambda_i\to\kk_f,\lambda_f)=\frac{2\pi
m}{\hbar^2k_i}\left|\langle\lambda_f,\kk_f|V|\lambda_i,
\kk_i\rangle\right|^2\delta(E_{ik}+E_{i\lambda}-E_{fk}+E_{f\lambda})\frac{L^6d^3k_f}{(2\pi)^3}.
\end{equation}
Taking into account that $E_{fk}=\hbar^2k_f^2/2m$, $d^3k_f=mk_fdE_fd\Omega_f/\hbar^2$,
one obtains double differential scattering cross section
\begin{equation}\label{g2a}
\frac{d^2\sigma}{dE_fd\Omega_f}(\kk_i,\lambda_i\to\kk_f,\lambda_f)=\frac{m^2k_f}{\hbar^4k_i}
\left|\langle\lambda_f,\kk_f|V|\lambda_i,
\kk_i\rangle\right|^2\delta(E_{ik}+E_{i\lambda}-E_{fk}+E_{f\lambda})\frac{L^6}{(2\pi)^2}.
\end{equation}
To compare with experimentally measured value one averages
(\ref{g2a}) over initial states and sums over final states of the
scatterer, getting
$$\frac{d^2\sigma}{dE_fd\Omega_f}(\kk_i\to\kk_f,{\cal
P})=$$
\begin{equation}\label{gg3a}
\sum\limits_{\lambda_i,\lambda_f} {\cal
P}(\lambda_i)\frac{m^2k_f}{\hbar^4k_i}
\left|\langle\lambda_f,\kk_f|V|\lambda_i,
\kk_i\rangle\right|^2\delta(E_{ik}+E_{i\lambda}-E_{fk}+E_{f\lambda})\frac{L^6}{(2\pi)^2},
\end{equation}
where  ${\cal P}(\lambda_i)$ is the probability of scatterer to be initially in the state
$|\lambda_i\rangle$.
\end{enumerate}
In the case, when
$$V=\frac{\hbar^2}{2m}4\pi b\delta(\rr_n-\rr_s),$$
where $\delta$-function depends on neutron and scatterer positions
$\rr_n$, $\rr_s$ respectively, the cross section (\ref{gg3a})
becomes
\begin{equation}\label{g3a}
\frac{d^2\sigma}{dE_fd\Omega_f}(\kk_i\to\kk_f,{\cal P})
=\sum\limits_{\lambda_i,\lambda_f} {\cal
P}(\lambda_i)\frac{k_f}{k_i}
\left|\langle\lambda_f|\exp(i\ka\rr_s)|\lambda_i\rangle\right|^2
\delta(E_{ik}+E_{i\lambda}-E_{fk}+E_{f\lambda}),
\end{equation}
where $\ka=\kk_i-\kk_f$ is momentum transferred to the scatterer.

One of contradictions here is that the probability of scattering
per unit time (\ref{wpg1a}) does not depend on time, so integral
of $dW_F$ over time is senseless, whereas according definition it
should be less than unity.

Two other dubious points are: introduction of the superfluous
volume $L^3$ and of flux density for a {\bf single} particle.

The SST has also another drawback --- one never calculates
scattering amplitude in it. However this amplitude can be
important. With it one would be able to find coherent amplitude,
which is averaged over initial states $|\lambda_i\rangle$ of the
scatterer. This amplitude gives coherent scattering cross section,
which some times could carry an additional information about
scattering process.

\subsection{The Fundamental Scattering Theory (FST)}

In FST (see, for example~\cite{gbw,tay}), which was developed
mainly to justify SST, the scattering process is divided into
three stages: infinite past, where incident particle is free;
present, where the neutron interacts with a scatterer; and
infinite future, where scattered neutron is again a free particle.
To work within the Hilbert space of normalized states, the
incident free neutron is represented by a wave packet
$|\phi\rangle$ with Fourier expansion:
\begin{equation}
|\phi\rangle\equiv|\phi(\kk,s)\rangle=\int\limits_{}^{}d^3pa(\kk-\p,s)|\p\rangle,
\label{wpg1}
\end{equation}
where parameter $s$ characterizes dimension of the wave packet,
$\kk$ is momentum of the packet, $|\p\rangle$ is a state
corresponding to plane wave, $\langle\rr|\p\rangle=\exp(i\p\rr)$,
in coordinate space, and $a(\p)$ are numerical Fourier
coefficients. The dynamics of the packet $|\phi\rangle$ is
determined by free hamiltonian $H_0$:
\begin{equation}
|\phi(t)\rangle=\exp(-iH_0t)|\phi\rangle. \label{wpg0}
\end{equation}
After scattering the wave function becomes
\begin{equation}
|\chi\rangle=\int\limits_{}^{}d^3p'|\p'\rangle
\int\limits_{}^{}d^3p\langle\p'|\hat S|\p\rangle a(\kk-\p,s),
\label{wpg24}
\end{equation}
where $\hat S$ is scattering matrix. And dimensionless probability
of scattering is defined as
\begin{equation}
dw(\kk\to\p')=d^3p'|\langle\p'||\chi\rangle|^2=
d^3p'\left|\int\limits_{}^{}d^3p\langle\p'|\hat S|\p\rangle
a(\kk-\p,s)\right|^2. \label{tayl0}
\end{equation}
This last formula is absolutely wrong. It means that scattering
transforms the wave packet into a plane wave. Such a process is
impossible because it completely violates unitarity. The state
$|\phi\rangle$ is normalized and $|\p\rangle$ is not normalizable.

From unitarity of the $S$-matrix it follows that norm of the wave
function is preserved. So the final wave function $|\chi\rangle$
after scattering is ether a wave packet or a superposition of wave
packets
\begin{equation}
|\chi\rangle=\int\limits_{}^{}d^3k'b(\kk\to\kk')|\phi(\kk',s')\rangle
\label{wpg9}
\end{equation}
with some new $s'$ if the parameter $s$ depends on $k$. Then
$b(\kk\to\kk')=\langle\phi(\kk',s')|\hat S|\phi(\kk,s)\rangle$
defines the amplitude of transition probabi\-li\-ty of the wave
packet state $|\phi(\kk,s)\rangle$ with momentum $\kk$ into the
wave packet state $|\phi(\kk',s')\rangle$ with momentum $\kk'$.

In fact, in the books~\cite{gbw,tay} and others only scattering of
plane waves is considered, and the initial wave packet defines
only spectrum of plane waves in the incident beam. However plane
waves scattering gives only probability, and averaging over
initial spectrum of plane waves does not change the situation.

In FST one also introduces the finite volume $L^3$ with an
arbitrary large $L$ for justifi\-ca\-ti\-on of SST. Below we show
that it is possible to avoid this artificial step. We use
fundamental scattering theory with plane waves (FPWST) and
calculate scattering pro\-ba\-bi\-li\-ty of neutron in monatomic
gas. This probability, as is shown in introduction, is ambiguous,
however, on one side, there is a way to find the standard
expression for cross section, and on the other side, knowledge of
the ambiguity helps to find the way to fight it.

The authors of~\cite{gbw,tay} and other books, make one special
step to get cross section from probability, however this step
contains a connivance that there is no scattering, if the
scatterer does not cross the neutron wave packet. In linear
scattering theory it is false, as we shall see in section 5.
Nevertheless we shall accept this point, because otherwise it is
absolutely impossible to get a cross section from probability. We
shall accept such a connivance as an implicit introduction of
nonlinearity into neutron-atom interaction.

\section{Direct calculation of neutron scattering from a mo\-na\-to\-mic gas}

Here we apply the principles of FST to plane wave neutron-atom
scattering and show that the result is ambiguous. We also show how
one can obtain the known result of SST.

\subsection{The scattering amplitude (\protect\ref{intr})\label{sec}}

When we consider neutron scattering from a monatomic gas, we must treat the
neutron and atom of the gas in the same way, it means that we need the same Schr\"odinger
equation for both particles
\begin{equation}
\left[i\frac{\partial}{\partial
t}+\frac{\Delta_n}{2}+\frac{\mu\Delta_a}{2}-
\frac{1}{2}u(\rr_n-\rr_a,t)\right]\psi(\rr_n,\rr_a,t)=0,
\label{2aa2}
\end{equation}
where potential $u$ is
\begin{equation}
u(\rr,t)=4\pi b\delta(\rr_n-\rr_a), \label{wpg1d}
\end{equation}
$\rr_n$, $\rr_a$, $m$, $M$ are coordinates and masses of the
neutron and atom respectively, $\mu=m/M$, and the time $t$
contains factor $\hbar/m$.

The Green function of equation (\ref{2aa2}) with excluded
interaction is
\begin{equation}
G(\rr_n-\rr'_n,\rr_a-\rr'_a,t-t')=\int\limits_{}^{}
\frac{\exp(i\kk_f(\rr_n-\rr'_n)+\p_f(\rr_a-\rr'_a)-i\omega(t-t'))}
{E_{fk}+E_{fp}-\omega-i\epsilon}\frac{d^3k_fd^3p_fd\omega}{(2\pi)^7},
\label{1b}
\end{equation}
where $E_{fk}=k_f^2/2$, $E_{fp}=\mu p_f^2/2$. It can be easily
checked by substitution that $G$ satisfies the equation
\begin{equation}
\left[i\frac{\partial}{\partial
t}+\frac{\Delta_n}{2}+\frac{\mu\Delta_a}{2}
\right]G(\rr_n-\rr'_n,\rr_a-\rr'_a,t-t')=-\delta(\rr_n-\rr_n')\delta(\rr_a-\rr_a')\delta(t-t').
\label{grenf}
\end{equation}

The scattered part of the wave function is
$$\delta\psi=
\frac{2\pi b}{(2\pi)^7}\int\limits_{}^{} \frac{d^3k_fd^3p_fd\omega
d^3r'_nd^3r'_adt'}{E_{fk}+E_{fp}-\omega-
i\epsilon}\exp(i\kk_f(\rr_n-\rr'_n)+i\p_f(\rr_a-\rr'_a)-
i\omega(t-t'))\times$$
$$\delta(\rr'_n-\rr'_a)\exp(i\kk_i\rr'_n+i\p_i\rr'_a-i
(E_{ik}+E_{ip})t')=$$
\begin{equation}
=\frac{b}{(2\pi)^2}\int\limits_{}^{}
\frac{d^3k_fd^3p_f\delta(\kk_f+\p_f-\kk_i-\p_i)}{E_{fk}+ E_{fp}-
E_{ik}-E_{ip}-i\epsilon} \exp(i\kk_f\rr_n+i\p_f\rr_a-
i(E_{ik}+E_{ip})t), \label{a1}
\end{equation}
where $\exp(i\kk_i\rr_n-iE_{ik}t)$, $\exp(i\p_i\rr_a-iE_{ip}t)$
describe incident plane waves of the neutron and atom respectively
with their energies $E_{ik}=k_i^2/2$, and $E_{ip}=\mu p_i^2/2$.

The wave function (\ref{a1}) after integration over $\p_f$ and
$\kk_f$ can be found in the final form
\begin{equation}
\delta\psi=\frac{b}{1+\mu}
\frac{1}{|\rr_n-\rr_a|}\exp\left(\frac{i}{1+\mu}[q|\rr_n-\rr_a|+i\PP(\mu\rr_n-\rr_a)]-iEt\right),
\label{aa11}
\end{equation}
where $q=|\kk_i-\mu\p_i|$ is relative speed of particles,
$\PP=\p_i+\kk_i$ is their total momentum, and $E=E_{ik}+E_{ip}$ is
their total energy. This wave function has a very clear physical
interpretation, however it is not an asymptotical one.

To get the asymptotical wave function we represented (\ref{a1}) as
a superposition of plane waves related to final states of the
neutron, $\exp(i\kk_f\rr_n-iE_{fk}t)$, and atom,
$\exp(i\p_f\rr_a-iE_{fp}t)$:
\begin{equation}\label{ab11}
\delta\psi=\int\limits_{}^{}\widetilde{F}'(\kk_i,\p_i\to\kk_f,\p_f,t)
d^3k_fd^3p_f\exp(i\kk_f\rr_n+i\p_f\rr_a-iE_{fk}t-iE_{fp}t),
\end{equation}
where
\begin{equation}\label{ac11}
\widetilde{F}'(\kk_i,\p_i\to\kk_f,\p_f,t)= \frac{b}{(2\pi)^2}
\frac{\delta(\kk_i+\p_f-\kk_i-\p_i)}{E_{fk}+E_{fp}-
E_{ik}-E_{ip}-i\epsilon} \exp(i(E_{fk}+E_{fp}- E_{ik}-E_{ip})t).
\end{equation}

With the relation (\ref{mt1}) we find in the limit $t\to\infty$
that the probability amplitude for the particle to leave in the
state $\kk_f$, and for the atom to leave in the state $\p_f$ is:
\begin{equation}\label{ad11}
\lim_{t\to\infty}\widetilde{F}'(\kk_i,\p_i\to\kk_f,\p_f,t)d^3k_fd^3p_f=
d^3k_fd^3p_f\frac{ib}{2\pi} \delta(\kk_f+\p_f-\kk_i-\p_i)
\delta(E_{fk}+E_{fp}-E_{ik}-E_{ip}),
\end{equation}
which coincides with (\ref{intr}).

\subsection{How to get the standard cross section}

Now we can show that with this amplitude it is possible to get the
standard cross section. For that we transform to CM reference
frame and obtain the probability scattering amplitude
(\ref{intr9}):
\begin{equation}\label{fbcm}
dF(\kk_i,\Omega_c,\p_i)=\frac{ibq}{2\pi(1+\mu)^2}d\Omega_{c},
\end{equation}
from which with the help of front area $A$ and (\ref{intr10}) we
obtain the cross section in the CM frame
\begin{equation}\label{bcm}
d\sigma(\kk_i,\Omega_c,\p_i)=A\left|\frac{ibq}{2\pi(1+\mu)^2}\right|^2
d\Omega_{c}
\end{equation}
for neutron scattering on an atom with the initial momentum $\p_i$.

Transformation to the laboratory frame according to (\ref{intr7b})
gives
\begin{equation}\label{intr11}
d\tilde\sigma_l(\kk_i\to\kk_f,\p_i)=Aq\left|\frac{ib}{2\pi(1+\mu)}\right|^2
d^3k_f\delta(E_R-\omega+\mu\p_i\ka).
\end{equation}
The number of collisions per unit time in a gas sample is
$$\nu(\kk_i\to\kk_f,\p_i)=qN_0d\tilde\sigma_l(\kk_i\to\kk_f,\p_i),$$
where $q$ is relative velocity. Since the interaction time with
the whole sample of width $d$ is $t=d/k_i$, then the total
probability of a single neutron scattering in the sample is
$$W(\kk_i\to\kk_f,T)=N_0\frac{d}{k_i}\int\limits_{}^{}d^3p_iq\sigma(\kk_i\to\kk_f,\p_i){\cal M}
\left(\mu\frac{p_i^2}{2T}\right),$$
where ${\cal M}(\mu p_i^2/2T)$ is Maxwellian distribution of $\p_i$:
$${\cal M}\left(\frac{\mu p_i^2}{2T}\right)=\left(\frac{\mu}{2\pi T}\right)^{3/2}
\exp\left(-\mu\frac{p_i^2}{2T}\right).$$
If we accept
\begin{equation}\label{area}
A=\frac{[2\pi(1+\mu)]^2}{q^2},
\end{equation}
and define cross section as $\sigma=W/N_0d$, then we obtain the
standard scattering cross section in a monatomic gas
\begin{equation}
d\sigma(\kk_i\to\kk_f,T)=\frac{|b|^2d^3k_f}{k_i\kappa\sqrt{2\pi\mu
T}} \exp\left(-\frac{(E_R-\omega)^2}{4E_RT}\right). \label{mxw2}
\end{equation}
It is necessary to point out that definition (\ref{area}) of the
parameter $A$ means that the wave packet contracts with increase
of the relative energy, and because of that the wave optics at low
energies transforms to geometrical one at high energies.

\paragraph{The total cross section for an atom at rest.}

If instead of Maxwellian distribution we substitute ${\cal M}(\p_i)=\delta(\p_i)$,
we immediately obtain the total cross section
\begin{equation}\label{bcmm1}
\sigma(\kk_i\to\kk_f,\p_i=0)=4\pi\left|\frac{b}{1+\mu}\right|^2,
\end{equation}
which also coincides with the standard one.

\subsubsection{Direct calculations in LF}

All the standard results were obtained with the help of CM
reference frame. If we repeat the same with scattering probability
(\ref{intr6}) obtained directly in the laboratory reference frame
we find a divergency when averaging with Maxwellian distribution
and an additional factor
$$Q(\mu)=\frac{8}{3}\mu^2+\frac{(1-\mu^2)^{3/2}}{\mu}
\arctan\left(\frac{\mu}{\sqrt{1-\mu^2}}\right)$$ in (\ref{bcmm1})
for scattering from an atom at rest.

\section{About wave packets}

Introduction of the parameter $A$ means that instead of plane
waves we deal with wave packets. We cannot deduce $A$ from a
theory, however we can study its properties. We also need to learn
how to describe scattering in terms of wave packets. Here we
consider elastic scattering of a wave packet from a fixed center,
and find at first sight unexpected though finally well
understandable result that probability of scattering does not
depend on position of the scattering center. However first of all
we need to discuss the types of the wave packets.

\subsection{Three types of the wave packets}

In general all the wave packets can be represented by their
Fourier expansion
\begin{equation}
\psi(\rr,\kk,t,s)=G(s|\rr-\kk
t|)\exp(i\kk\rr-i\omega(k,s)t)=\int\limits_{}^{}d^3p\,a(\p,\kk,s)
\exp[i\p\rr-i\omega(\p,\kk,s)t], \label{q4q4}
\end{equation}
where parameter $s$ determines the width of the packet like in
(\ref{d1}), $a(\p,\kk,s)$ and $\omega(\p,\kk,s)$ are functions of
invariant variables $\kk^2$, $\p^2$ and $\kk\p$.

The primary wave packet describes a free incident particle. Its
Fourier expansion contains plane waves $\exp(i\p\rr)$, which
satisfy the equation
\begin{equation}
[\Delta+p^2]\exp(i\p\rr)=0. \label{e1}
\end{equation}
All the packets are representable in the form (\ref{q4q4}), and
they differ by the Fourier coefficients $a(\p,\kk,s)$ and
dispersion $\omega(\p,\kk,s)$. We consider three types of the wave
packets and discuss which one is the most appropriate for
description of particles.

\subsubsection{The Gaussian wave packet}

The most popular in the literature is the Gaussian wave packet
\begin{equation}
\psi_G(\rr,\kk,t,s)=\left(\frac{s}{\sqrt\pi(1+its^2)}\right)^{3/2}
e^{i\kk\rr-ik^2t/2}\exp\left(-\frac{s^2[\rr-\kk
t]^2}{2[1+its^2]}\right). \label{g1}
\end{equation}
This packet is normalized to unity, satisfies the free
Schr\"odinger equation, but spreads in time. Because of this
spreading its form in space does not completely coincides with
(\ref{q4q4}).

Its Fourier components are
\begin{equation}
a_G(\p,\kk,s)= \left(\frac{1}{2\pi
s\sqrt\pi}\right)^{3/2}\exp(-(\kk-\p)^2/2s^2),
\quad\omega_G(\p,\kk,s)=p^2/2, \label{q4q5}
\end{equation}
where $s$ is the width in momentum space. The spectrum of wave
vectors $\p$ is spherically symmetrical with respect to the
central point $\p=\kk$ and decays away from it according to
Gaussian law.

The cross area of this packet moving in $z$ direction can be
defined as
\begin{equation}
A_G= \int\limits_{}^{}\pi\rho^2d^3r|\psi_G(\rr,\kk,t,s)|^2=
\int\limits_{}^{}\rho^2d^2\rho \frac{s^2}{1+t^2s^4}
\exp\left(-\frac{s^2\rho^2}{1+t^2s^4}\right)=
\pi\frac{1+t^2s^4}{s^2}, \label{wpg43}
\end{equation}
where $\rho=\sqrt{x^2+y^2}$.

\subsubsection{Nonsingular de Broglie wave packet}

It is known that there are no nonspreading normalizable wave
packets, which satisfy the free Schr\"odinger equation. However
nonnormalizable wave packets do exist. As an example we can
demonstrate nonsingular de Broglie wave packet~\cite{bro}
\begin{equation}
\psi_{ns}(\rr,\kk,t,s)=\exp(i\kk\rr-i\omega(k,s) t)j_0(s|\rr-\vv
t|), \label{q4q2}
\end{equation}
in which $\omega(k,s)=k^2/2+s^2/2$ and $\vv=\kk$  in units
$\hbar/m=1$. The packet (\ref{q4q2}) is a spherical Bessel
function $j_0(sr)\exp(-is^2t/2)$, which center is moving with the
speed $\vv$. This packet satisfy the free Schr\"odinger equation.
Its Fourier components are
\begin{equation}
a_{ns}(\p,\kk,s)\propto \delta((\kk-\p)^2-s^2),\quad
\omega_{ns}(\p,\kk,s)=p^2/2, \label{q4q6}
\end{equation}
and spectrum of $\p$ is a sphere of radius $s$ in momentum space
with center at the point $\p=\kk$. Since it is not normalizable,
its front area is infinite, and such a wave packet is not a good
candidate for description of free neutrons.

\subsubsection{The singular de Broglie wave packet}

The singular de Broglie wave packet~\cite{bro}
\begin{eqnarray}
\psi_{dB}(\rr,\kk,t,s)= C\exp(i\kk\rr-i\omega(k,s)
t)\frac{\exp(-s|\rr-\vv t|)} {|\rr-\vv t|}, \label{exc3}
\end{eqnarray}
is normalizable one with normalization constant $C=\sqrt{s/2\pi}$
defined by
\begin{eqnarray}
\displaystyle \int d^3r \vert {\mit \psi}_{dB} (\rr,\kk,t,s)
\vert^2 = 1. \label{exc4}
\end{eqnarray}
The parameter $s$ is the width of the packet in momentum space and
reciprocal width in coordinate space, $\vv\equiv\kk$ is the packet
speed, and $\omega(k,s)=(k^2-s^2)/2$. We see that $\omega(k,s)$ is
less than kinetic energy by the term $s^2/2$, which can be thought
of as bound energy of the packet.

The singular de Broglie wave packet satisfies inhomogeneous
Schr\"odinger equation
\begin{equation}
\left[i\frac{\partial}{\partial
t}+\frac{\Delta}{2}\right]\psi_{dB}(\rr,\vv,t,s)= -2\pi
Ce^{i(v^2+s^2)t/2}\delta(\rr-\vv t), \label{ab1}
\end{equation}
with right hand side being zero everywhere except one point.

The Fourier coefficients of the singular de Broglie wave packet
are
\begin{equation}
a_{dB}(\p,\kk,s)=
\sqrt{\frac{s}{2\pi}}\frac{4\pi}{(2\pi)^3}\frac{1}{(\p-\kk)^2+s^2}.
\label{d1d}
\end{equation}
and
\begin{equation}
\omega_{dB}(\p,\kk,s)=[2\kk\p-k^2+s^2]/2=[p^2-(\kk-\p)^2-s^2]/2.
\label{d1cc}
\end{equation}
The spectrum of wave vectors $\p$ is spherically symmetrical with
respect to the central point $\p=\kk$ and decays away from it
according to Lorenzian law with width $s$.

The Fourier coefficients (\ref{d1d}) and frequency (\ref{d1cc})
become identical to those of spherical wave
\begin{equation}
\exp(-ik^2t/2)\frac{\exp(ikr)}{r}=
\frac{4\pi}{(2\pi)^3}\int\limits_{}^{}\exp(i\p\rr)
\frac{\exp(-ik^2t/2)\,d^3p}{p^2-k^2-i\epsilon}, \label{sph33}
\end{equation}
after substitution $\kk\to0$ and $s\to -ik$, i.e.
$\psi_{dB}(\rr,0,t,-ik)= C\exp(ikr-ik^2t/2)/r$.

The front area of the singular de Broglie wave packet, moving in
$z$ direction can be defined as
\begin{equation}
A_{dB}=\int \pi\rho^2d^3r|\psi_{dB}(\rr,\vv,t,s)|^2=
\frac{s}{2\pi}\int\limits_{0}^{\infty}2dz\pi d\rho^2\pi\rho^2
\frac{\exp(-2s\sqrt{\rho^2+z^2})}{\rho^2+z^2}, \label{wpg44}
\end{equation}
where $\rho=\sqrt{x^2+y^2}$. After change of variables
$\xi=z/\rho$ we get
\begin{equation}
A_{dB}=2\pi s\int\limits_{0}^{\infty}d\xi d\rho\rho^2
\frac{\exp(-2s\rho\sqrt{1+\xi^2})}{1+\xi^2}=
\frac{\pi}{2s^2}\int\limits_{0}^{\infty}\frac{d\xi}{(1+\xi^2)^{5/2}}=
\frac{\pi}{3s^2}. \label{wpg45}
\end{equation}

\subsubsection{Genesis of the singular de Broglie wave packet}

The singular de Broglie wave packet descends from the spherical
wave. Indeed, let's consider the spherical wave with energy
$q^2/2$:
\begin{equation}
\psi(r,t,q)= \exp(-iq^2t/2)\frac{\exp(iqr)}{r}. \label{q5q}
\end{equation}
This wave satisfies inhomogeneous Schr\"odinger equation
\begin{equation}
\left[i\frac{\partial}{\partial
t}+\frac{\Delta}{2}\right]\psi(r,t,q)= -2\pi
\exp(-iq^2t/2)\delta(\rr). \label{ab2a}
\end{equation}
The right hand side describes the center radiating the spherical
wave. If we change to reference frame moving with the speed
$\vv=\kk$ then we must perform the following transformation of the
function $\psi$:
\begin{equation}
\psi(r,t,q)\to \Psi(\rr,\kk,t,q)=
\exp(i\kk\rr-ik^2t/2-iq^2t/2)\frac{\exp(iq|\rr-\kk t|)}{|\rr-\kk
t|}. \label{q4q}
\end{equation}
The transformed function is the spherical wave around moving
center. It satisfies the equation
\begin{equation}
\left[i\frac{\partial}{\partial t}+\frac{\Delta}{2}\right]\Psi=
-2\pi \exp(i[k^2-q^2]t/2)\delta(\rr-\kk t). \label{ab2b}
\end{equation}
If the energy of the wave (\ref{q5q}) is negative: $q^2=-s^2$,
i.e. the wave (\ref{q5q}) describes a bound state around the
center, then (\ref{q4q}) becomes
\begin{equation}
\Psi(\rr,\kk,t,is)=
\exp(i\kk\rr-ik^2t/2+is^2t/2)\frac{\exp(-s|\rr-\kk t|)}{|\rr-\kk
t|}. \label{q4q1}
\end{equation}
With normalization constant $C$ expression (\ref{q4q1}) becomes
identical to (\ref{exc3}). Thus the singular de Broglie wave
packet is the spherical Hankel function of imaginary argument
moving with the speed $\vv$.

\subsubsection{Genesis of the nonsingular de Broglie wave packet}

The nonsingular de Broglie wave packet is obtained by
transformation to the moving reference frame of the nonsingular
spherical wave
$$j_0(qr)\exp(-iq^2t/2),$$
which satisfies the homogeneous Schr\"odinger equation. This way
we can construct a lot of nonsingular wave packets corresponding to
different angular momenta $l$.

\subsection{Scattering of wave packets from a fixed center}

We see that the proof of validity of SST is not perfect because of
unacceptable definition of scattering probability, according to
which a wave packet after scattering transforms to plane waves,
though according to unitarity it should remain a wave packet. Now
we look at a wave packet elastic scattering from a fixed center.
We take the wave packet not as a preparation of a particle in some
state, but as an intimate property of the particle, which means
that after scattering the particle is the same packet as before
it.

The wave packet (\ref{q4q4}) relates to a free particle. In the
presence of a potential $u(\rr)/2$ the plane wave components
$\exp(i\p\rr)$ should be replaced by wave functions
$\psi_{\p}(\rr)$, which are solutions of the equation
\begin{equation}
[\Delta+p^2-u(\rr)]\psi_p(\rr)=0
\label{e2x}
\end{equation}
containing $\exp(i\p\rr)$ as the incident wave. Substitution of
$\psi_p(\rr)$ into (\ref{q4q4}) transforms it to
\begin{equation}
\psi(\rr,\kk,t,s)=\int\limits_{}^{}d^3p\,a(\p,\kk,s)\psi_p(\rr)\exp[-i\omega(\p,\kk,s)t].
\label{q4q4x}
\end{equation}

Now we have to find asymptotical form of (\ref{q4q4x}). For that
we substitute asymptotical form of $\psi_p(\rr)$. For incident
wave $\exp(i\p\rr)$ asymptotical wave function after scattering on
a fixed center with an impact parameter $\ro$ is a superposition
of plane waves:
\begin{equation}
\psi_p(\rr)\Rightarrow\exp(i\p\ro)\int\limits_{}^{}d\Omega
F'(p,\Omega) \exp(i\p_\Omega[\rr-\ro]), \label{e2y}
\end{equation}
where $F'(p,\Omega)$ is the probability amplitude of a plane wave
with wave vector $\p$ to be transformed to the plane wave with
wave vector $\p_\Omega$ pointing into direction $\OO$ in the
element of solid angle $d\Omega$. This amplitude for isotropic
scattering is $F'(p,\Omega)=bp/2\pi$. Dependence on $p$ is an
irritating moment, however, since the spectrum of wave packets has
a sharp peak at $p=k$, we can approximate $F'(p,\Omega)$ by
$bk/2\pi$, having in mind that corrections to this value is of the
order $s/k$, where $s$ is the packet width in the momentum space.
This correction is small, when $s$ is small, i.e. area $A$ of the
packet is large.

The vector $\p_\Omega$ in (\ref{e2y}) is of length $p$, but
it is turned by angle $\Omega$ from $\p$.
Substitution of (\ref{e2y}) into (\ref{q4q4}) for $\exp(i\p\rr)$
transforms (\ref{q4q4}) to the form
\begin{equation}
\psi(\rr,\kk,t,s)=\int\limits_{}^{}d^3p\,a(\p,\kk,s)
\exp(i\p\ro)d\Omega F'(k,\Omega)
\exp[i\p_\Omega[\rr-\ro]-i\omega(\p,\kk,s)t]. \label{q4qq}
\end{equation}
Since $a(\p,\kk,s)$, $\p\ro$ and $\omega(\p,\kk,s)$ are invariant
with respect to rotation, we can replace them with
$a(\p_\Omega,\kk_\Omega)$, $\p_\Omega\ro_{\Omega}$ and
$\omega(\p_\Omega,\kk_\Omega,s)$. After that we can transform
integration variable $\p\to\p_\Omega$, and drop the index $\Omega$
of $\p$. As a result we transform (\ref{q4qq}) to the form
\begin{equation}
\psi(\rr,\kk,t,s)=\int\limits_{}^{}d^3p\,a(\p,\kk_\Omega,s)
\exp(i\p\ro_{\Omega})d\Omega F'(k,\Omega)
\exp[i\p[\rr-\ro]-i\omega(\p,\kk_\Omega,s)t], \label{q4qq1}
\end{equation}
which can be represented as
\begin{equation}
\psi(\rr,\kk,t,s)= \int\limits_{}^{}d\Omega F'
(k,\Omega)\psi_0(\rr-\ro+\ro_{\Omega},\kk_\Omega,t,s),
\label{q4qq2}
\end{equation}
where $\psi_0$ denotes the wave packet of the the same form as that of the
incident particle.

We see that the packet as a whole is scattered with probability
$dw(\kk\to\kk_\Omega)=|F'(k,\Omega)|^2d\Omega=|bk/2\pi|d\Omega$,
which, surprisingly, has no dependence on impact parameter $\ro$
as in the case of plane waves. It shows that scattering of wave
packets is almost the same as that of plane waves.

The independence of scattering amplitude on position of scatterer
is well understand\-able in linear wave mechanics. Indeed, the
wave packet is a superposition of plane waves, which exist in the
whole space. They cancel each other outside the packet, but they
exist, and because they are scattered independently of each other,
the whole packet's scattering does not depend on position of the
scatterer.

To get a cross section we need an additional hypothesis which
restricts scattering to those cases, when the wave packet overlaps
the target position. This hypothesis is outside of the wave
mechanics and in the books~\cite{gbw,tay} it is accepted
implicitly. We can say that without this hypothesis the wave
mechanics is incomplete theory, i.e. it is insufficient to
describe scattering of particles. Introduction of this hypothesis
is equivalent to inclusion of nonlinearity into quantum mechanics.
When we write $\sigma=Aw_1$, we implicitly accept nonlinearity.

\subsection{Experimental investigation of the wave packet properties}

We cannot deduce $A$, but we can explore its properties. So we can
ask ourselves how large is $A$ and what is its dependence on $E$.

\subsubsection{Size of $A$}

At first sight $A$ should not be large. Indeed, if cross section
of scattering from a fixed center, or from a heavy nucleus
($\mu\approx0$), according to (\ref{bcm}) is $4\pi
A|b|^2q^2/(2\pi)^2$, then, to get usual cross section $4\pi|b|^2$,
we must put $A=(2\pi/q)^2=\lambda^2$, where $\lambda$ is wave
length of the neutron motion with respect to the center.

However, if $A$ is small, then the total reflection of thermal
neutrons becomes impossible. Indeed, the wave packet (\ref{q4q4})
contains plain waves with different wave vectors $\p$. Probability
of wave packet reflection is determined by reflected wave function
\begin{equation}
\psi_r(\rr,\kk,t,s)= \int\limits_{}^{}d^3p\,R(p_\bot)a(\p,\kk,s)
\exp[i\p\rr-i\omega(\p,\kk,s)t], \label{q4q4q}
\end{equation}
where
$$R(p_\bot)=\frac{p_\bot-\sqrt{p_\bot^2-u}}{p_\bot+\sqrt{p_\bot^2-u}}.$$
The width of the spectrum of $\p$ is proportional to $s$. If
$A\propto1/s^2$ is small, then $s$ is large. If $s$ is large, the
probability to find $p_\bot^2>u$ when $k_\bot^2<u$ is also large,
so $\int d^3r |\psi_r(\rr,\kk,t\to\infty,s)|^2<1$. To get large
reflection we need small $s$, so for $k_\bot^2<u$ probability to
find plane wave components with $p_\bot^2<u$ is small. In that
case $\mu=1-\int d^3r |\psi_r(\rr,\kk,t\to\infty,s)|^2\propto s
\ll1$. In particular, if $k^2<u$, which is the case for ultracold
neutrons (UCN), $\mu\le3\cdot10^{-5}$, which means~\cite{uts} that
$s\approx3\cdot10^{-5}k$ (\ref{area1}).

However, if $s$ is small, than $A\propto1/s^2$ is large. Then to
get
$$4\pi\frac{A|b|^2q^2}{(2\pi)^2}=4\pi\frac{|b|^2}{9(2\pi)^2}\cdot10^{10}$$
to be equal to the measured cross section of the order 1 barn, we
need to take $b$ of the order of $10^{-15}$ cm. With such small
$b$ we obtain optical potential $u=4\pi N_0b$ too low to see UCN
storage, and, more over, again we cannot see the total reflection
of thermal neutrons. So we have a new contradiction: $A$ have to
be large, but it cannot be large. Later we shall see how to
resolve this contradiction.

In any case the value of $A$ can be explored experimentally in two
types of experiments: in measurement of transmission of thin films
when the grazing angle of incident neutrons is below critical
one~\cite{uts1}, and deviation of reflection angle from the
specular direction, when the grazing angle of incident neutrons is
a little bit above the critical one~\cite{gus}. Both of these
effects are proportional to $s$, as it follows from (\ref{q4q4q}).

\subsubsection{Energy dependence}

We have seen in section 4.2 that to get the standard cross section
for neutron-monatomic gas scattering we need to suppose that $A$
is proportional to $1/E$, where $E$ is the energy of relative
neutron-atom motion. If the area $A$ were constant then the cross
section calculated via CM reference frame would depend on gas
temperature $T$ proportionally~\cite{apoc} $T^{3/2}$. So, to
define whether $A$ does depend on energy or not, it is necessary
to measure temperature dependence of neutron transmission through
monatomic gas.

There is one publication~\cite{gen} on this matter. However it was
not the temperature dependence, which was measured there, but
energy dependence of neutrons transmission through noble gases.
The temperature dependence was deduced, because transmission
depends on dimensionless parameter $E/T$. So, it was desirable to
measure directly the temperature dependence. The experiment was
done with cold neutrons~\cite{igt} and it had shown, that cross
section grows with temperature proportionally to $T^{1/2}$, which
supports $A\propto1/q^2$ dependence.

This results contradicts to the experiment on subcritical
transmission of mirrors, when grazing angle of incident neutrons
is below critical one~\cite{uts1}. However, we believe that the
last experiment should be repeated to seek more carefully for
transmitted neutrons.

\section{The list of contradictions and their resolution}

\begin{enumerate}
\item
The SHT theory, which is used for description of elastic
scattering and two body scattering in CM reference frame
contradicts canonical quantum mechanics, because spherical waves
do not correspond to free particles after scattering.

This contradiction is resolved, when we find asymptotical form of
spherical waves in accordance with FST.
\item
SST obtains cross sections by introduction of scattering
probability per unit time (though it does not depend on time), and
flux density for a single particle.

It can be avoided, if instead of SST we use FST.
\item
FST pretends to prove validity of SST, however this prove is
inconsistent, because it violates unitarity. This violation is
related to transformation of initial wave packet into final plane
waves.

We resolve this contradiction replacing wave packet in the initial
state by a plane wave.
\item
However, then we obtain only probabilities and not cross sections.

This contradiction is resolved by introduction of wave packet and
its cross area $A$.
\item
To improve FST we consider both initial and final neutrons
described by the same wave packet. However we find that wave
packets scatter nearly the same as plane waves. The scattering
probability does not depend on impact parameter, so
introduc\-ti\-on of wave packets does not help to get cross
sections from probabilities.

Resolution of this contradiction requires introduction of
nonlinearity into interaction. We do not have such a theory so we
include $A$ by hands, and suppose that such inclusion implicitly
involves nonlinearity.
\item
When we consider the value of $A$ we find that it should be large
but cannot be large.

We resolve this contradiction simultaneously with the following
ambiguity.
\item
When we apply FST with initial and final plane waves we find that
calculation of neutron scattering from a moving atom gives
different results, when we find scattering probability directly in
LF reference frame, or via intermediate transformation to CM.

Below we show how this ambiguity can be resolved, and this
resolution simultaneous\-ly resolves the contradiction between
(\ref{area}) and (\ref{area1}).
\end{enumerate}

In the whole we must claim that there are no now a selfconsistent
scattering theory. However we are to proceed with that we have at
our hands.

\subsection{Quantization of scattering}

The ambiguity arises because of the continuous spectrum of
angular distribution
\begin{equation}
\psi_s=\int\limits_{2\pi}^{}F'(k,\Omega)d\Omega\exp(i\kk_\Omega\rr).
\label{wpg39n}
\end{equation}

No ambiguity would appear, if the angular distribution were
discrete. So we shall eliminate the ambiguity, if we replace the
integral in (\ref{wpg39n}) with the integral sum
\begin{equation}
\psi_s=\sum\limits_{j=1}^{N}F'(k,\Omega_j)\delta\Omega_j\exp(i\kk_{\Omega_j}\rr)=
\sum\limits_{j=1}^{N}\delta F_j\exp(i\kk_{\Omega_j}\rr),
\label{wpg39o}
\end{equation}
where discrete amplitudes $\delta
F_j=F'(k,\Omega_j)\delta\Omega_j$ are introduced. In such a
representation the probability of scattering into the angle
$\Omega_j$ is $|\delta F_j|^2$, and probability of scattering into
the angular interval covering $2n$ elements $j_1-n\le j\le j_1+n$
around some direction $\Omega_{j1}$, is
$$dw_n(\Omega_{j1})=\sum\limits_{j=j_1-n}^{j_1+n}|\delta F_j|^2\approx w'(\Omega_{j1})\Delta\Omega,$$
where
$$\Delta\Omega=\sum\limits_{j=j_1-n}^{j_1+n}\delta\Omega_j\approx2n\delta\Omega_{j1},\qquad
w'(\Omega_{j1})=|F'(k,\Omega_{j1})|^2\delta\Omega_{j1},$$ and we
supposed that $w'(\Omega_{j})$ are almost constant in the interval
$j_1-n\le j\le j_1+n$.

Transformation of the integral to the sum can be made with an
arbitrary choice of $\delta\Omega_j$ around every $\Omega_j$,
which is also an ambiguity. To resolve it we can make a step in
style of quantization, i.e. we can require all the amplitude
elements $F_j=F'(k,\Omega_j)\delta\Omega_j$ to be equal, which
means that we introduce a quantum of the area
$\int_{\delta\Omega}F'(k,\Omega)d\Omega$. In the CM reference
frame, where $F(k,\Omega)$ is constant, such a requirement means
an introduction of quantum of the angular interval or uncertainty
$\delta\Omega$. Below we discuss what value this quantum can be
of.

When we transform from one reference frame to another, we see a
deformation of both $F'(k,\Omega)$ and $\delta\Omega$, however the
amplitude elements $\delta F(\Omega)=F'(k,\Omega)\delta\Omega$ and
the number of such elements in the whole integral
$\int_{4\pi}F'(k,\Omega)d\Omega$  remain the same. Therefore, if
we take some amount of amplitude elements in the center of mass
reference frame, square every element, and after that transform to
the laboratory reference frame, then we obtain the same amount of
the squared elements in the laboratory frame, and they will be
confined in some angular interval. If we transform from center of
mass to laboratory frame without squaring the amplitude elements,
and square them only after the transformation, we obtain the same
number of squared elements in the same angular interval as before.
Which means that our result is completely invariant under Galilean
transformation. Therefore we have the full right to use center of
mass reference frame without a danger to get an ambiguous result
after changing of the reference frame.

\subsection{A value of the scattering angle quantum}

According to (\ref{fbcm}) we have
\begin{equation}\label{acmf2n}
F'(q,\Omega_{c})=\frac{ibq}{2\pi(1+\mu)^2}.
\end{equation}
Therefore the scattering amplitude element is
\begin{equation}\label{acmf2n1}
\delta F(\Omega_{c})=\frac{ibq}{2\pi(1+\mu)^2}\delta\Omega.
\end{equation}
Probability of scattering into some interval of solid angle
$\Delta\Omega=n\delta\Omega$ is
\begin{equation}\label{acmf2n2}
\Delta w(\Omega_{c})=n|\delta
F(\Omega_{c})|^2=\left(\left|\frac{bq}{2\pi(1+\mu)^2}\right|^2
\delta\Omega\right)\Delta\Omega,
\end{equation}
the differential scattering cross section becomes
\begin{equation}\label{acmf2n3}
\frac{d\sigma(\Omega)}{d\Omega}=\frac{\Delta
w}{\Delta\Omega}A=\left|
\frac{bq}{2\pi(1+\mu)^2}\right|^2A\delta\Omega,
\end{equation}
and we immediately obtain that to get
$d\sigma/d\Omega=|b|^2/(1+\mu)^2$, we have to require
\begin{equation}\label{acmf2n4}
\left|\frac{q}{2\pi(1+\mu)}\right|^2A\delta\Omega=1.
\end{equation}
It means that the angular quantum $\delta\Omega$ correlates with
the wave packet cross area $A$! This correlation is very important
because it resolves the contradiction of two results: (\ref{area})
and (\ref{area1}). Since, without doubt, the interval
$\delta\Omega$ is small, we must have $A$ to be large. So the
contradiction between (\ref{area}) and (\ref{area1}) is resolved
in favor to (\ref{area1}).

\section{Conclusion}

Simple consideration of a scattering process shows a contradiction
inherent in the SHT. On one side we use plane waves as eigenstates
of a particle, and on the other side describe scattered particles
with spherical waves, which are not even solutions of the free
Schr\"odinger equation. Rigorous approach, which resolves this
contradiction, creates another one. With this approach we can
calculate only dimensionless probabilities of scattering, while
for interpretation of experiments we need cross sections with
dimensions of area. To resolve this second contradiction and to
get cross sections instead of dimen\-si\-on\-less probabilities,
we have to introduce some front area $A$ of the wave function for
the incident particle. Thus we arrive at wave packets, and must
describe scattering process with wave packets instead of plane
waves. However, while doing that we arrive at the next
contradiction: the wave packets scatter similarly to plane waves,
and scattering does not depend on whether the scatterer crosses
the wave packet or not. This contradiction is a result of
linearity of the quantum mechanics. To resolve this contradiction
we must introduce a nonlinearity at least into interaction term to
assure that scattering takes place only, when the scatterer is
inside the front area of the wave packet. However, even with this
hypothesis in mind we have another contradiction: the wave packet
have to be large, but cannot be large. If we abandon wave packet
and calculate scattering probability with plane waves, we find
that the probability of scattering is an ambiguous value, which
depends on a way of calculation. To resolve this ambiguity we are
to quantize the process of scattering. This quantization leads to
replacement of angular integrals with discrete integral sums. It
is amazing that resolution of the ambiguity simultaneously
resolves the contradiction related to the value of area $A$
pointed out in (\ref{area}) and (\ref{area1}). The last result
proves that some truth is contained indeed in our investigations.

As for SST it is not a theory, because it is not justified by FST,
but only a set of rules how to calculate cross sections. It can be
used by everyone, who is satisfied with these rules,
notwithstanding that they are not well justified. We hope that our
research will help to understand better the merits of SST , of
quantum mechanics as a whole, and will give an impetus for serch
of a more consistent scattering theory.

\section*{Acknowledgement}
The author is very grateful to J-F. Bloch, prof. of Institute
Francais Politechnigue de Grenoble, for his interest to the work
and assistance, to Dr. E.P.Shabalin of JINR (Dubna) for his
support, to F.Gareev and V.Furman for their encouragement, to
V.Nikolenko for his help, to D.Koltun, prof of Rochester
University, and to Dr. A.Shabad from Lebedev Physics institute of
RF academy of science for their useful discussions.

\section*{History of submissions}

\subsection{Submission to Phys. Lett. A}
On 09 March 2005 I submitted the paper to Phys. Lett. A and only
on 17 July I received the following reply:

On 09-Mar-2004 you submitted a new paper number
PLA-ignatovi.AT.jinr.ru-20040309/1 to Physics Letters A entitled:

'Dramatic problems of Scattering Theory'

We regret to inform you that your paper has not been accepted for
publication. Comments from the editor are:

I apologise for the delay in writing to you about your paper.

Despite my considerable efforts I have been unable to elicit any
referee reports on your article and at this stage I do not think I
am going to receive any reviews. In the circumstances and to save
you losing further time I believe it would be in your interests if
you submitted your article to another journal.

Thank you for submitting your work to our journal and I wish you
luck in publishing it elsewhere. Yours sincerely,

J.P. Vigier Editor Physics Letters A

\subsection{Submission to Phys. Rev. A}

On August 3 I submitted the paper to Phys. Rev. A with the
following covering letter:

Please, I would not like this manuscript will go to hands of the
Associated editor Lee A. Collins. I prefer to get referee reports
and not rejection from him.

I wrote this letter, because I had an experience, that Lee A.
Collins rejected my papers without considerations.

\subsubsection{Referee report on September 10}

Dear Dr. Ignatovich,

The above manuscript has been reviewed by one of our referees.
Comments from the report are enclosed.

We regret that in view of these comments we cannot accept the
paper for publication in the Physical Review.

Yours sincerely,

Stephen J. Smith Associate Editor Physical Review A Email:
pra@ridge.aps.org Fax: 631-591-4141 http://pra.aps.org/

----------------------------------------------------------------------

Report of the Referee -- AH10019/Ignatovich

----------------------------------------------------------------------

This paper makes the startling claim that there are fundamental
flaws in the standard quantum scattering theory that was developed
by Moller, Jauch, and others and is described in Refs 4 and 5 and
many others texts (for example, {\it Scattering Theory of Waves
and Particles} by Roger Newton). Claims of this sort must be taken
seriously. But they must also be supported by unusually good
evidence, and, in my view, such evidence is definitely not present
here. Therefore, it is my opinion that this paper should not be
published in anything like its present form.

I shall not discuss all of the points raised in the paper, but
three will, I hope, suffice. Consider first the claim in
connection with Eqs. (60) and (61) that standard scattering theory
violates unitarity. The function $a({\bf k -p})$ is the
momentum-space wave function of the incoming state
$|\psi_{in}\rangle$. The function $\chi({\bf p}) = \langle{\bf
p}|\chi \rangle$, with $|\chi\rangle$ as defined in Eq. (60), is
the momentum-space wave function of the corresponding outgoing
state $|\psi_{out}\rangle = S|\psi_{in}\rangle$. If
$|\psi_{in}\rangle$ is normalized, then so is
$|\psi_{out}\rangle$, since $S$ is unitary. Equation (61) is
simply the elementary proposition that the absolute value squared
of the momentum-space wave function gives the probability density
that a measurement of momentum will give an answer in the
neighorhood of $\bf p$. It in no way implies that the scattering
process has produced a transition from the normalized
$|\psi_{in}\rangle$ to the improper state $|\bf p\rangle$. Rather,
the scattering process has produced the normalized state
$|\psi_{out}\rangle$ with a well-defined probability for its
momentum to be found in the neighborhood of any chosen point $\bf
p$. The discussion of Eq. (61) here and elsewhere seems to show a
failure to understand the most basic principles of quantum
mechanics.

Consider next the observation that the wave function of Eq. (28)
is not a solution of the free Schr\"{o}dinger equation everywhere.
This observation is, of course, correct, but is in no way in
conflict with standard scattering theory.  The form (28) is only
valid in the asymptotic  region, far outside the target potential.
Thus it is only expected to satisfy the free Schr\"{o}dinger
equation in this region.  Its behavior near $r = 0$ is completely
irrelevant.  Thus the presence of the delta function $\delta(r)$
in Eq. (29) is, in no way, the flaw that the author seems to
believe it to be.

Finally, the claim on page 29 that the scattering probability is
independent of the position of the target  violates both common
sense and observed fact.  Any rigorous definition of the
scattering cross section requires an integration over impact
paramenters, and this integration converges only if the scattering
probabiliy goes to zero as the impact parameter gets large.  This
is true in the standard scattering theory of the standard
references.  If it is false in the author's version of scattering
theory, this only proves that his version is undeniably
unsatisfactory.

This paper should not be published.
\subsubsection{My reply on September 14}

Dear referee,

1)  You write "consider Eqs. (60) and (61). The function $a({\bf k
-p})$ is the momentum-space wave function of the incoming state
$|\psi_{in}\rangle$."

No, it is an amplitude of momentum state component $\exp(i{\bf
pr})$ in the composition of the initial state. If you mean the
same but call it a momentum wave function - then it is only
semantic difference between us.

 "The function $\chi({\bf p}) = \langle{\bf p}|\chi \rangle$,
 with $|\chi\rangle$ as defined in Eq. (60), is the
 momentum-space wave function of the corresponding outgoing
 state $|\psi_{out}\rangle = S|\psi_{in}\rangle$."

My reply is the same as about incident neutrons. It is only
amplitude of the plane wave component. If you mean the same, let's
not discuss semantics.

 "If $|\psi_{in}\rangle$ is normalized, then so is
 $|\psi_{out}\rangle$, since $S$ is unitary."

I agree, and I told about it in my paper just in the second
paragraph after eq. (61).

"Equation (61) is simply the elementary proposition that the
absolute value squared of the momentum-space wave function gives
the probability density that a measurement of momentum will give
an answer in the neighborhood of $\bf p$."

I agree and ask you: what is the wave function, which describes
the final state with momentum $\bf p$, what is its dependence on
coordinates and time?

"It in no way implies that the scattering process has produced a
transition from the normalized $|\psi_{in}\rangle$ to the improper
state $|\bf p\rangle$."

I don't agree that it does not imply, I agree that it should not,
because $|\bf p\rangle$ is improper state.

 "Rather, the scattering process has produced the normalized
 state $|\psi_{out}\rangle$ with a well-defined probability
 for its momentum to be found in the neighborhood of any chosen point $\bf p$."

I agree that it produced the normalized state, however my
question: What the scattering process means: transition from
improper state $\bf p'$  i.e. $\langle{\bf p'}|\psi_{in}\rangle$
to improper state $\langle{\bf p}|\psi_{out} \rangle$ or initial
state has a well definite momentum $\bf k$? Then what is the wave
function with momentum $\bf p$ after scattering? Look, please,
$S$-matrix transforms plane waves into superposition of plane
waves.

Why don't you comment my eq. (62) and text below it? If particles
are described by normalized wave functions the scattered wave
function should have momentum density not $|\langle{\bf
p}|\psi_{out}\rangle|^2$, but $\langle{\phi({\bf
k'})}|\psi_{out}\rangle$, where $\bf k'$ momentum of scattered
particle with normalized outgoing wave function ${\phi({\bf
k'})}$. This is my argument. Why do you ignore it?

2)  You write: "The form (28) is only valid in the asymptotic
region, far outside the target potential.  Thus it is only
expected to satisfy the free Schr\"{o}dinger equation in this
region.  Its behavior near $r = 0$ is completely irrelevant."

I agree. However I don't speak about behavior near $r=0$.

"Thus the presence of the delta function $\delta(r)$ in Eq. (29)
is, in no way, the flaw that the author seems to believe it to
be."

My arguments: Even bound state wave function satisfies free
Schrodinger equation outside the potential well. You do not
consider it as scattering wave function, because it has
exponentially decaying asymptotic, not $1/r$. However the function
$1/r$ also contains exponentially decaying waves. Therefore to be
self consistent you must neglect exponentially decaying waves
there too. Then you obtain real asymptotic --- the superposition
of plane waves. Why don't you discuss eq-s. (32), (33)? For
clarity I added few words after eq. (29).

3)  You write: "the claim on page 29 that the scattering
probability is independent of the position of the target  violates
both common sense and observed fact."

My reply. I agree about common sense and observed facts. However I
discuss the theory, and show that the standard theory is unable to
provide agreement with common sense and observed facts.

"Any rigorous definition of the scattering cross section requires
an integration over impact parameters, this integration converges
only if the scattering probability goes to zero as the impact
parameter gets large."

My reply. Oh I am grateful to you for this remark. It means you
agree that we have no right to consider plane waves, because for
plane waves the impact parameter has no sense.

 "This is true in the standard scattering theory of the standard references."

My reply: It is not true, because nobody considered scattering of
wave packets into wave packets. Or you can give me a reference?

  "If it is false in the author's version of scattering theory,
  this only proves that his version is undeniably unsatisfactory."

My reply. I agree that such result is unsatisfactory. For me
myself it was a tragedy, when I discovered it. And I submitted
this paper in order to show: friends, look, what happens with
standard approach. (It is standard approach, not my version.) You
reject result, and didn't show where mathematics is wrong in my
consideration of the wave packet scattering. Take into account,
please, that I am the first, who tried to do it, and why don't you
discuss my arguments after eq. (99)? If something is not clear,
give me to know, please. I'll try to improve.

Dear referee, I appeal not to a student, who had just accumulated
his knowledge, but to a researcher, who dares to doubt. I
understand that you have no time to read the paper carefully, but
I ask you then to take a responsibility before killing a paper. I
am ready to discuss all my points openly. I hope that you will be
one of those, who will be interested enough to spare more time for
discussion.

Yours sincerely, the author.

\subsubsection{Reply of non anonymous referee on November 1 and my reply on
November 02}

Dear Dr. Ignatovich,

This is in reference to your appeal on the above mentioned paper.
We enclose the report of our Editorial Board member Peter Mohr
which sustains the decision to reject.

Stephen J. Smith Associate Editor Physical Review A Email:
pra@ridge.aps.org Fax: 631-591-4141 http://pra.aps.org/

----------------------------------------------------------------------

Report of the Editorial Board Member -- AH10019/Ignatovich

----------------------------------------------------------------------

{\bf For not to repeat the report, I included my replies in it in
bold face style}

The manuscript AH10019, "Contradictions of quantum scattering
theory," by Vladimir K. Ignatovich has been rejected by Physical
Review A, and the author has requested reconsideration.  The
decision was based on the position of the referee that the
evidence presented in the manuscript is not sufficient to show
that conventional scattering theory has fundamental flaws, as
claimed by the author.  The referee singled out three examples of
points made in the manuscript that were not convincing in this
regard.  The author has attempted to show that the comments made
by the referee are not valid.

The first point concerns Eq. (61) and the statements made in the
manuscript: "This last formula is absolutely wrong.  It means that
scattering transforms the wave packet into a plane wave."  The
referee does not accept this statement, because the formula does
not mean that the scattered wave function is a plane wave.  The
referee correctly points out that Eq. (61) represents the
probability density that the scattered particle will have a
particular momentum and that it does not imply that the final
state is a plane wave.  This is the conventional and correct
quantum mechanical interpretation of such an equation.  The
subsequent discussion in the manuscript does not support the
author's claim, as suggested in his resubmission letter.

{\bf My reply. Before to find momentum density of scattered
particle, you should define the wave function of the particle.
Your disagreement with my arguments shows how many unclear notions
quantum mechanics contains.}

The second point concerns Eq. (29).  In the manuscript it is
stated that "the spherical wave does not describe a free
particle." In fact, outside of the region containing the origin,
the spherical wave does describe a free particle, as pointed out
by the referee, and the claim by the author is not valid.  The
author goes on to suggest that his claim supported by the fact
that bound-state wave functions outside of the potential decay
exponentially, however this analogy is not convincing. In fact,
the solution in Eq. (29) is a free solution outside of the
potential for either a scattering state or a bound state.  The
only difference is that $k$ is imaginary in the case of a bound
state.

{\bf My reply. Why don't you comment my equations (32) and (33)? I
tell just about imaginary component of $k$. The waves, which have
imaginary component in some direction must be excluded from
asymptotic.}

The third point made by the referee is that the conclusion by the
author that the scattering of a wave packet is independent of the
impact parameter is problematic.  The author claims that this is a
problem with conventional scattering theory, however it appears
more likely that the author's interpretation of scattering theory
is at fault.

{\bf My reply. Your words ``more likely'' show that you express an
opinion of a common man but not of a scientist!}

The source of the problem is indicated in the statement in the
manuscript: "The independence of the scattering amplitude on
position of scatterer is well understandable in linear wave
mechanics. Indeed, the wave packet is a superposition of plane
waves, which exist in the whole space. They cancel each other
outside the packet, but they exist, and because they are scattered
independently of each other, the whole packets scattering does not
depend on position of the scatterer." The last statement is not
correct and overlooks the fact that if the initial state is a wave
packet, then the plane wave components will scatter individually
in such a way that their coherent superposition will destructively
interfere to give a small scattering probability if the impact
parameter of the initial wave packet is large.

{\bf My reply. I thought the same, and discovered that it is
wrong! Read the paper, please.}

Evidently, the calculation in the manuscript makes approximations
that lose this information.  However, it is not appropriate for
the referee to correct the manuscript line by line in order to
show how to fix the problem.

{\bf My reply. Then you have no right to claim that calculations
are wrong!}

In addition to the valid criticisms of the referee, it should be
noted that the description of scattering in terms of wave packets
is not a new result.  For example, the subject is discussed in the
article "Scattering of Wave Packets," by Eyvind H. Wichmann,
American Journal of Physics, 33, 20 (1965).  The abstract of that
article notes that: "The aim of this paper is mainly pedagogical,
and the final results obtained are old and well established."

{\bf My reply. I thank you very much for the reference. I rushed
into library, grasped the journal, buried myself in it. and was
greatly disappointed. There are no wave packet at all, except
words. May be it is a special pedagogical trick of Am.J.Phys.?}

Based on the above points, the recommendation not to publish this
paper is upheld here.

Dr.Peter J. Mohr Editorial Board Member Physical Review A

{\bf Dear Dr. Mohr, it is very difficult to discuss the paper
point by point with a man, who does not want to listen arguments.
I was surprised that you did not understand my simplest arguments
about exponential decay of some part of spherical wave asymptotic
(eq. (32) and (33)). Do you understand that the
$\sqrt{k^2-p_\|^2}$ becomes imaginary, when $p_\|^2>k^2$? If you
do not understand that, how can you understand my mathematics
about wave packet scattering?!

I know that your letter does not suppose discussion, so I do my
last step and appeal to Editor-in Chief of APS.

The Author}

\subsubsection{My last reply and appellation to Editor-In-Chief of
APS on November 2}

 Dear Editor, below (here above) I wrote my reply to Dr.
Mohr. I consider his report as incompetent one. Forward, please,
my letter with my reply and my entire file to the Editor-In-Chief
of the American Physical Society.

The letter to the Editor-In-Chief

Dear, Editor-In-Chief, you already have one of my appellations on
your table. It is on the paper about experiment on measurement of
temperature dependence of neutron-helium scattering, submitted to
Physical Review Letters. Referees of that paper complained that
they are unable to see my full paper on contradictions. I tried
hard to publish this paper in Physical Review A. Look, please, how
difficult it is, and how incompetent are referee reports. Both
referee even do not understand that the square root
$\sqrt{k^2-p^2}$ becomes imaginary, when $p^2>k^2$! I am ready to
defend all my results openly! Give me a chance, please, to do it.
I also ask your permission to put all my discussions with referees
at the end of my paper. I suppose it is the only way to punish
irresponsible referees! Yours sincerely, Vladimir Ignatovich.

\subsubsection{The letter of an editor on 8-th November and my
reply to it}

Dear Dr. Ignatovich,

This is with regard to your message of 02 November regarding your
manuscript referenced above. Your message appers to reflect a
misunderstanding on your part of the editorial process.

Your paper was submitted to Physical Review A on 03 August 2005.
This Journal rejected your manuscript on 09September 05, on the
basis of an incontrovertibly negative referee report. You appealed
the rejection to the Editorial Board; the appeal was adjudicated
by Board Member Dr. Peter Mohr who denied your appeal. Now you
apparently ask that the rejection of your appeasl by the Board be
further appealed to the Editor-in-Chief of the American Physical
Society. However, the closing sentence(s) of the Board
adjudication state that "Under the revised Editorial Policies of
the Physical Review (copy enclosed), this completes the scientific
review of your paper."

While you may request that the case be reviewed by the
Editor-in-Chief of the APS (see enclosed Editorial Policies). This
request should be addressed to the Editor, who will forward the
entire file to the Editor-in-Chief of the APS.Suvch an appeal must
be based on the fairness of tehprocedures followed, and must not
be a request for another scientific review. The question to be
answered in this review is: Did the paper receive a fair hearing?
The decision of the Editor-in-Chief concludes the consideration of
the manuscript by the American Physical Society.

Yours sincerely,

Bernd Crasemann Editor Physical Review A

{\bf My reply}

Dear Dr.Bernd Crasemann, thank you for reply. May I ask you, are
you just that editor to whom I am to ask to forward my appeal to
Editor-in Chief. If yes, please, forward this letter and all my
file to him.

May I ask you? What does it mean "fair hearing?" If I claim that
2x2=4, and referee reports that he learned from text books that
2x2=5, and Dr. Mohr repeats that he believes the same, is it fair
hearing.

May I ask all your editorial board to answer a simplest question:
does the integral in my formula (32) contain exponentially
decaying waves or not? Dr Mohr replies "not", and I don't believe
he is a scientist. You should check his diploma.

I shall appeal to all the Physical Society of America. I shall
publish all my correspondence on this paper to demonstrate how
many slaves are in science!

Forward, please, all my file to the Editor-In-Chief. I would like
to see, where can I find a real scientist with whom it is possible
to discuss the science and not ideology. Sincerely, V.K.I.

P.S. If I am in error, and you are not the proper editor, may I
ask you to forward this letter and my appeal to the proper Editor?
In any case, I shall be grateful for your reply and advice.

\subsubsection{Reply of Editor-in-chief Martin Blume of 13 Dec. 05 sent by postal mail}

I have reviewed the file concerning this manuscript which was
submitted to Physical Review A. The scientific review of your
paper is the responsibility of the editor of Physical Review A,
and resulted in the decision to reject your paper. The
Editor-in-Chief must assure that the procedures of our journals
have been followed responsibly and fairly in arriving at this
decision.

The insulting language that you have used in your correspondence
does nothing to advance the cause of your manuscript, nor has it
any place in scientific discourse. We will not respond to any
further correspondence on this matter.

On considering all aspects of this file I have concluded that our
procedures have in fact been appropriately followed and that your
paper received a fair review. Accordingly, I must uphold the
decision of the Editors.


\begin{thebibliography}{99}
\bibitem{igt}
Temperature dependence of neutron transmission by He-4. To be
published.
\end{thebibliography}
\end{document}